# The optical calibration system for the CTAO North Large Sized Telescope

M. Iori[1,2], L. Recchia[2], F. Ferrarotto[2], E. Budai[4], F. Di Pierro[3], C. Vigorito[3,4]
on behalf of the CTAO-LST Project

[2] *INFN Roma1 Università La Sapienza P.zza A. Moro 5, 00185 Rome Italy*
[3] *INFN Sezione di Torino, Via P. Giuria 1, 10125 Torino, Italy*
[4] *Dipartimento di Fisica - Università degli Studi di Torino, Via Pietro Giuria 1 - 10125 Torino, Italy*

In 2018, the first prototype of the optical calibration system was installed in the Large Sized Telescopes, LST-1 at the Observatorio del Roque de los Muchachos, ORM, in La Palma, Canary Islands. We have almost completed the construction of the three remaining ones that will be mounted by 2026 on LST-2-3-4. The main feature of this system is to ensure a stable and constant photon flux over time with the characteristics of a signal produced by Cherenkov light when a high-energy gamma ray crosses the atmosphere. Test results that show the performance of the optical calibration system have been recently published. In this article, we present in detail the system used to monitor the photon flux.

## 1. Introduction

Within 2026, the CTAO Consortium will complete the construction and installation of all four planned Large Sized Telescopes (LST-1-4) at Observatorio de Roque de Los Muchachos (ORM) in La Palma, Canary Islands, and we will finish equipping all with the optical calibration system (CaliBox). The LST cameras require a precise and regular consistent calibration. The CaliBoxes are designed to fulfill the requirements needed for the calibration of the cameras, including monitoring the photon flux to guarantee the quality of the system for laser stability, uniform illumination, and intensity range. The results of the tests on the performance of the calibration system were presented in [1]. In this article, we show the details of the system for monitoring the photons that are sent to the camera. The need to monitor the photons with high precision is due to the fact that the components of the optical system, such as the laser, the filters, and the diffuser can deteriorate over a long period of data taking. For this reason, we have installed a silicon photomultiplier sensor, SiPM, at one exit of the diffuser that monitors the outgoing flux. In this paper, we show the dedicated electronics to translate the SiPM analog signal into a digital value and the specific calibration of the electronic components necessary to evaluate the number of photons that are provided for each event.

## 1.1 Description of the device

The apparatus has been designed to ensure that the photon flux emitted by the laser is not altered by variations related to humidity, dust and temperature. For this reason, the optical components, the 355 nm pulsed laser, the Ulbricht sphere, and the optical filters are inside two aluminium boxes connected to each other and filled with nitrogen at atmospheric pressure. The electronic components, the relays, the processors that manage the pulsed flux of the 400 ps laser pulsed beam, the combination of filters used to attenuate the photon flux, and the sensors that monitor the photon flux of the optical system are outside the boxes containing the optics. Depending on the combination of filters, the photon flux at the entrance of the Ulbricht sphere varies from 1 to $10^5$ photons. Fig. 1 shows the external view of the CaliBox and its block diagram. Table 1 summarizes the performance of the device required by the technical project [2] to achieve the high quality of the data. Fig. 2 shows where the CaliBox is installed in the center of the dish.

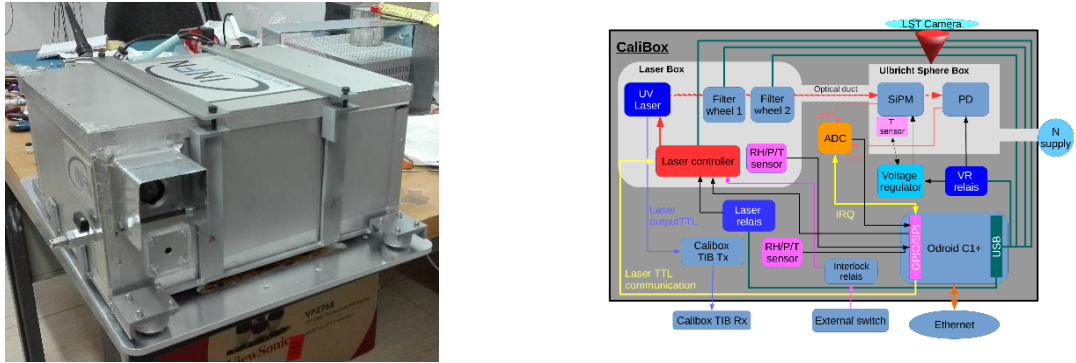

Figure 1: Left: External view of the INFN CaliBox; Right: the CaliBox block diagram [1].

**TABLE 1**
**Features of the optical calibration system for CTAO North LST-1-4**

| | |
|---|---|
| Laser Q-switching | 355 nm - 400 ps pulse width - 1-2 kHz - 1 µJ |
| Laser stability (a) | <1% |
| Attenuation of laser beam | UV ND filter $10^{-1}$ to $10^{-4}$ |
| Filter combinations | 35 |
| Laser pulse width at exit diffuser (a) | 2.8 ± 0.03 ns |
| Photon flux range (a) | 1 to $10^5$ |
| Uniformity of illumination at 5 m from device (a) | <1% |
| Monitoring photons precision, G(T) (b) | 3.0x$10^{-3}$ |
| RH inside optical box filled with N 1 atm | 20% |
| System control processors | Odroid-C1, Arduino |

(a) Ref [1], (b) this paper

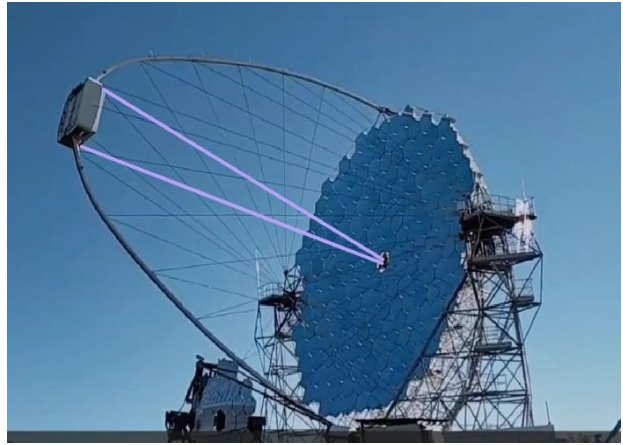

Figure 2: The LST-1 telescope. The purple lines show the diffused UV beam emitted by the CaliBox.

## 2- Estimation of the photon flux variability

A sample of 1000 events per point was taken, using a 100 Hz laser pulse frequency, at 10 minutes intervals for nine hours to reproduce the same data taking conditions of a typical night shift. The average integrated pulse signal versus time is stable within 1.0 ± 0.2% (rms over mean). Fig.3 shows the Gaussian fit of the averaged integral values over nine hours and the Gaussian fit of the averaged integral values over 100 s with 10 s step to show both long-term and short-term stability.

The peak-to peak stability representing the range of output power variation within nine hours, results to be equal to 1.2 %. This factor considers the laser output power fluctuation related to the temperature drift, pump power fluctuation and the crystal temperature change. In the measurement done in laboratory the ambient temperature shows there is no relation between the temperature variation in the laboratory and the beam laser intensity thanks to the fact that the optical system is inside a hermetically sealed box [1].

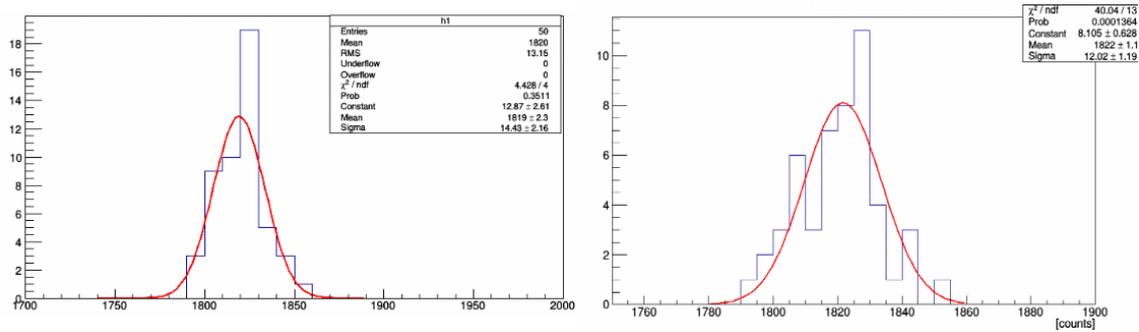

Figure 3. Left: The Gaussian fit of the averaged integral values over nine hours. Right: The Gaussian fit of the averaged integral values over 100 s with 10 s step.

## 3. Monitor of the photon flux

To monitor the photons, a Silicon photomultiplier is used, whose bias voltage is varied as a function of the temperature in order to maintain a constant gain. A CAEN A7585D module controlled by I2C Arduino varies the bias voltage as a function of the temperature according to a calibration made in laboratory. The A7585D is a high voltage regulator, VR, designed for SiPM bias. It has a built-in temperature compensation controller with a programmable coefficient. The VR can provide up to 10 mA and the output voltage can be regulated between 20 V and 85 V with a resolution of 1 mV. The gain, G, as shown in Fig. 4b, depends on the temperature because the lattice vibrations become stronger and this fact leads to an attenuation of the avalanche. The relation between the breakdown voltage, $V_{bd}$, and the temperature is linear and it can be written as

$$V_{bd}=V_{bd0}[1+ \beta(T-T_0)] \qquad (1)$$

where $V_{bd0}$ is the breakdown voltage at $T_0$ and $\beta$ is the $V_{bd}$ variation as function of the temperature. Hence, in order to keep the gain constant, the overvoltage $\Delta V = V_{bias} - V_{bd}$ must be tuned following the linear dependence shown in equation (1).

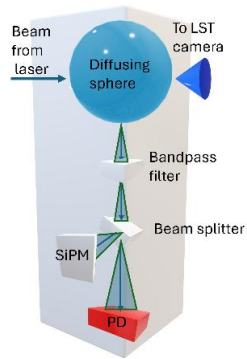

a)

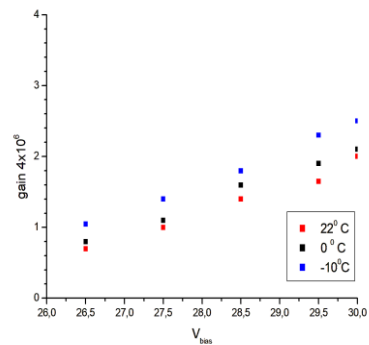

b)

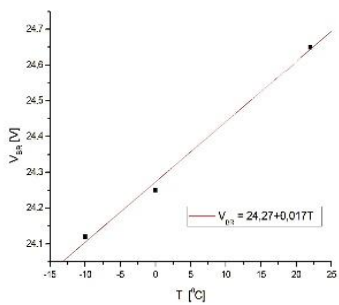

c)

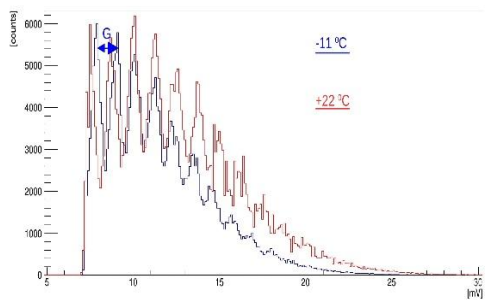

d)

Figure 4: a: Schematic design of the optical components that shows the Ulbricht sphere (black) and the position of the SiPM used to measure the photon flux; b: Gain versus $V_{bias}$ for different temperature values; c: dependence of the breakdown voltage versus the temperature; d: Typical charge distribution of SiPM illuminated by pulsed LED 400 nm light. It shows a constant gain in the interval T= 22 $^0$C (red) to -11 $^0$C (blue).

The correction function to maintain the SiPM gain constant was obtained by performing a dedicated test in the INFN Rome1 Laboratory using a climatic chamber in a temperature range from -11 to 22 °C and a pulsed blue 400 nm LED. The LED was located outside the climatic chamber to avoid being influenced by temperature changes and was connected to the sensor through a light guide. By conducting multiple measurements in the range from -11 to 22 °C, we have defined a function of the bias voltage as a function of the increase in Celsius degrees to maintain the gain constant. For the 3x3 mm$^2$ SensL MicroFC-30335 SiPM model [4] the compensation factor measured to maintain a constant gain β is 0.017 ± 0.001 V/°C. Fig. 4d demonstrates that using this parameterization, the gain remains constant in the range from -11 to 22 °C within $Tx3.10^{-3}$, and it is equal to $4.3x10^6$. This value includes the amplifier gain of the SiPM readout circuit corresponding to a factor about 10.

## 4- Evaluation of photon number sent to the camera

To evaluate the number of photons sent to the camera, the signal is translated from analogue to digital. The signal detected by the sensor with specific wheel combinations is integrated by a 12-bit ADC, triggered by the laser signal. Due to the SiPM signal duration of 50 ns, it is transformed using the peak-hold technique, designed and built at INFN Roma1 Electronics Lab, which monitors the signal and retains its peak value as output. Fig. 5 shows the SiPM signal and the peak-hold output with τ ~ 200 μs. A TTL signal generated by the ODROID pulses the laser and at the same time generates an IRQ signal to set the ADC readout. The precision of the ADC measurement is defined by the CS time position in the exponential signal generated by the peak-hold circuit. The measured average jitter of CS is about 100 *μs*. As shown in [1], by stepping through several filter combinations, we find a linear dependence between the variance and the ADC counts as expected by a Poisson distribution. The linear relation between the peak height of the SiPM signal measured and the ADC permits to evaluate the p.e. per ADC channel.

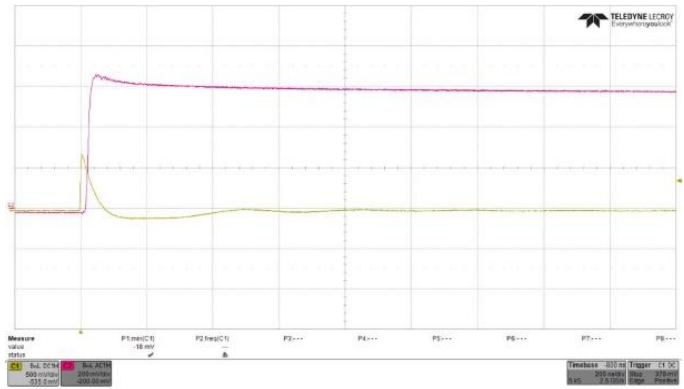

figure 5: Sipm signal (yellow) and the peak-hold output (red) The scope setup is 200 ns /div and 500 mV/div for the SiPM signal and 200 mV/div for the peak-hold output signal.

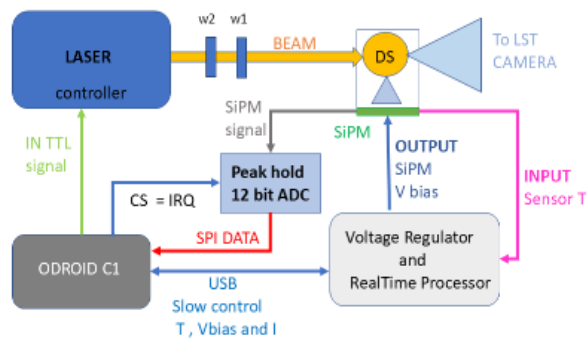

Figure 6: Block diagram of the system to monitor the photon flux sent to the camera with all details. In the figure DS = diffusing sphere. The figure shows all relevant electronic components and all buses and signal paths used to control all of them. The ODROID starts the laser via a TTL signal sent via GPIO; the ADC uses a dedicated peak and hold circuitry and a delayed TTL IRQ signal from the ODROID to measure the SiPM signal charge. The Voltage regulator senses the temperature near the SiPM and adjusts, according to a calibrated linear function, the bias Voltage of the SiPM to maintain the gain constant with changing temperatures.

**Conclusions**

An optical calibration system for calibrating the Large Sized Telescopes LST-1-4 has been designed to ensure stability in photon flux measurements by controlling the parameters that can alter it. The results from the laboratory tests show the calibration system satisfies the requirements of the CTAO project that requires a stability of the photon flux at 2%, a dynamic range of 100-400 photons per pulse as well as the uniformity of the illumination of the camera of 2% or less, [1]. We have shown the technique to monitor the photon flux obtained reading it at the exit of the Ulbricht sphere by using a SiPM sensor. The parametrization to adjust the $V_{bias}$ as function of the temperature and to get the gain constant it is stable at $Tx3.10^{-3}$.